\documentclass[onecolumn,showpacs,preprintnumbers,amsmath,amssymb,amsthm,mathrsfs]{revtex4}
\usepackage{graphicx}% Include figure files
\usepackage{dcolumn}% Align table columns on decimal point
\usepackage{bm}% bold math
\begin{document}

\title{Oscillatory behavior of the in-medium interparticle potential in hot gauge system with scalar bound states}
\author{Hui Liu}
 \email{liuhui@iopp.ccnu.edu.cn}
\author{Defu Hou}
 \email{hdf@iopp.ccnu.edu.cn}
\author{Jiarong Li}
 \email{ljr@iopp.ccnu.edu.cn}
\affiliation{%
Institute of Particle Physics, Central China Normal University,
Wuhan(430079), P.R.China
}%
\date{\today}% It is always \today, today,

\begin{abstract}
We investigate the in-medium interparticle potential of hot gauge
system with bound states by employing the QED and scalar QED
coupling. At finite temperature an oscillatory behavior of the
potential has been found as well as its variation in terms of
different free parameters. We expect the competition among the
parameters will lead to an appropriate interparticle potential
which could be extended to discuss the fluid properties of QGP
with scalar bound states.
\end{abstract}
%{\bf Key Words:}


\maketitle

\section{Introduction}
When concerning a hot/dense plasma, Debye screening picture is
widely accepted where the in-medium screening potential is
exponentially damping with distance. The damping rate, namely the
so-called Debye mass, is the reciprocal of the screening length
which scales the effective interactive radius of in-medium
particles. At high temperature, the mean free path of particle is
much larger than the screening length, therefore such kind of
system can be treated as weakly coupled gas composed of
quasi-particles with effective radius. Nevertheless this
short-distance dominant Debye screening is not the whole story.
Due to the dissimilar analytic structures of the effective boson
propagator, the screening potential may behave distinctly at
different situations and approximations, in which large distance
oscillations become equally significant.

To make it more pellucid, we follow notations in Ref.\cite{Sivak}
and defined the in-medium interparticle potential $V(r)$ in linear
response framework as\cite{Kapusta1}
\begin{equation}\label{potential}
V(r)=\frac{Q_1Q_2}{4\pi^2r}\
\mbox{Im}\int^\infty_{-\infty}dq\frac{qe^{iqr}}{q^2+F(0,q)},
\end{equation}
where $r$ is the distance between two particles, $Q_1$ and $Q_2$
are their charges, $Q=(\omega, \bf{q})$ is the four-momentum
transfer with $q=|{\bf q}|$, and $F$ is the longitudinal boson
self-energy defined by $F=(Q^2/q^2) \Pi_{00}$. Obviously, the
behavior of $V(r)$ is totally determined by the analytic
structures of the denominator of the integrand on the complex
plane. The denominator and the so-called dielectric function are
almost the same except for a difference of $Q^2$. So we can simply
say that the form of dielectric function determines the
interactive potential.

Take the Friedel oscillations and Debye screening as examples. In
the extremely high density but zero temperature environment, the
explicit expressions of $F$ for nuclear
matter\cite{Alonso,Alonso2} and gauge plasma\cite{Kapusta1} have
been obtained by ignoring the fermion masses. Due to the existence
of sharp fermi surface, the dielectric function contains branch
cuts, which dominate the contribution of contour integration at
large distance and lead the interparticle potential oscillate
considerably instead of monotonic damping. This phenomenon induced
by the sharp fermi surface is named as Friedel
oscillations\cite{Walecka,Kapusta1,Alonso,Sivak}. However at
non-zero temperature the Friedel oscillations will be strongly
suppressed because of the smeared Fermi surface at finite
temperature. Fortunately, although the branch cut contribution is
suppressed at finite temperature, the pole contribution stands
out, resulting in the so-called Yukawa
oscillations\cite{Alonso,Sivak} which dominate in a wide range of
distance. On the contrary, at high temperature but zero chemical
potential, there is only pole but no branch cut contribution. At
hard thermal loop(HTL) approximation, the poles of the integrand
are on the imaginary axis\cite{LeBellac} which grants Debye
screening. While in a not very high temperature environment, the
HTL approximation might not be eligible and thus one may try to
give up this approximation and employ a complete one-loop
self-energy. However, the analytic structures of the self-energy
are much more complicated in one-loop order. Actually, the
position of the pole contains both imaginary and real parts on the
complex plane which can only be identified numerically. The
emergence of the real part of the pole means the interparticle
potential is not monotonically damping but oscillating instead.
For example, if there is a pole in the integrand of
Eq.(\ref{potential}), whose imaginary and real parts are defined
as $q_i$ and $q_r$ respectively, the potential can be obtained by
performing the contour integral in the upper plane and recast as
\begin{equation}\label{osc potential}
V(r)=\frac{Q_1Q_2}{\pi(a^2+b^2)}\frac{e^{-q_i r}}{r}\ [a\cos (q_r
r) +b\sin ( q_r r)],
\end{equation}
where $a$ and $b$ are defined as the real and imaginary parts of
the residue,
\begin{equation}\label{ab}
\left.\frac{(q^2+F(q))'}{q} \right|_{q=q_r+i q_i}=a+i b,
\end{equation}
with the prime denoting for $\partial/\partial q$. Considering the
real part of $F$ is an even function, we can only concern the
residue in the first quadrant and double it in the final result.
From mathematical points of view, when the pole is on the
imaginary axis, the oscillating factor in the blanket of
Eq.(\ref{osc potential}) vanishes as a result of $q_r=0$, which
leads to Debye screening potential. In other words, if the pole
deviates from the imaginary axis, there exit oscillations.

Besides the mathematical description of the oscillations, we have
an another potential motivation on the discussion of the
oscillatory potential. As we all know the quark-gluon plasma (QGP)
produced at relativistic heavy ion collider (RHIC) is almost
likely to be perfect fluid at temperatures in the order of several
times of critical temperature $T\sim (1-3)T_c$. The fascinating
low viscous mechanism has attracted much attention but still
challenges both experimentalists and theoreticians. One possible
way to understand it is to calculate the shear viscous
coefficient. In the weakly-coupled regime many publications such
as
Refs.\cite{Hosoya,Gavin,Danielewicz,Oertzen,Baym,Heisenberg,Arnold1,Arnold2,
Liu1,jeon1,wang,wang2,Thoma,defu,Aarts,Jeon,Carrington,Basagoiti,Liu2},
no matter from kinetics or Kubo formula, are overestimating the
shear viscosity extracted from experiments or lattice
results\cite{Shuryak,Teaney,Nakamura}. The failure prompts to
examine the picture of weak coupling and short-distance screening
potential, and consider the strong coupling and/or long-distance
correlation. Apart from the attempts in weak coupling, some
scientists proposed strongly-coupled plasma and the liquid state
theory\cite{Thoma2,Peshier} to explain the perfect fluid QGP. One
should keep in mind that the necessary condition to form a liquid
state is the interparticle potential is  strong enough in coupling
as well as  long enough in distance\cite{March,Egelstaff}.
Obviously Debye potential is not a good candidate because of its
short distance domination character. Therefore it makes sense to
study the oscillatory potential with respect to all kinds of
dynamical and thermodynamical parameters so as to make sure if it
is possible to produce a liquid state in such a hot environment.

Shuryak and Zahed\cite{Shuryak2} raised the idea of coloered
loosely binary bound states to explain the small viscosity at the
near critical temperatures. In their picture, quark and anti-quark
are bounded together, leaving a weak Coulomb potential as a
remnant for interacting with the colored particles, such as
quarks, gluons or other bound states in QGP. They argued the
possible existing low-lying resonances led to large "unitarity
limit" scattering cross section and naturally were expected to
play an important role in the transport coefficients. In this
paper, we adopt this picture of relativistic plasma with bound
states, attempting to study their effect on the in-medium
interparticle potential by using QED and scalar QED (sQED) as toy
models. Since we do not intend to describe the real
electromagnetic system but just quote their interaction forms to
see the qualitative properties of the in-medium interparticle
potential, so the temperature, the coupling strength and the
masses of fermion and scalar bound state, are free parameters in
the following calculation.

\section{QED and sQED self-energy}

\begin{figure}
%\begin{minipage}[t]{0.5\linewidth}
 %\centering
 \begin{center}
   \resizebox{10cm}{!}{\includegraphics{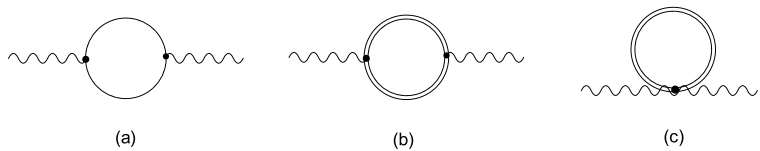}}
 \end{center}
\caption{One-loop boson self-energy. The solid line is for the
fermion and the double solid lines are for the scalar bound
state.\label{fig1}}
%\end{minipage}%
\end{figure}

In Eq.(\ref{potential}), one can see all the complication comes
from the longitudinal self-energy of gauge boson. In one-loop
order the photon self-energy is diagrammatically denoted by
Fig.1(a) in pure QED plasma and modified by adding
Figs.\ref{fig1}(b) and (c) according to sQED Feynman rules when
considering the existence of scalar bound states.
The explicit expression of $F$ for Fig.1(a) is composed of vacuum
and matter contributions, that are $F_{vac}^a$ and $F_{matt}^a$
respectively. The vacuum contribution can be obtained by
dimensional regulation and $\overline{MS}$ scheme. The result can
be found in many textbooks on Quantum Field Theory\cite{Peskin},
\begin{equation}\label{Favac}
F^{a}_{vac}(0,q)=- \frac{2\alpha}{\pi}q^2\int^1_0 dx x(1-x) \log
\left (\frac{m^2}{m^2+x(1-x)q^2}\right),
\end{equation}
where $\alpha=\frac{1}{137}$ is the fine structure constant of
QED, and $m$ is the mass of fermion. The matter contribution can
be obtained either in real time or imaginary time formalism which
reads\cite{Kapusta2}
\begin{equation}\label{Famatt}
F^{a}_{matt}(0,q)=-\frac{4\alpha}{\pi}\int^\infty_0 dp\
\frac{p^2}{\omega}\left[\frac{4\omega^2-q^2}{4p\ q}\log
\left(\frac{q-2p}{q+2p}\right)-1\right]n_f(\omega),
\end{equation}
where $\omega$ is the on-shell energy of fermion constrained to
$\omega=\sqrt{p^2+m^2}$ and $n_f(\omega)=(e^{\beta\omega}+1)^{-1}$
is the Fermi-Dirac distribution function with $\beta=1/T$.

The existing scalar bound states will "supplement" the
polarization of the gauge field and consequently remodel the
interaction among particles. We involve this effect by employing
scalar QED. The polarization and tadpole diagrams presented in
Figs.1(b) and (c) will be added to (a). In the real time formalism
$F$ can be written as\cite{Litim, Leupold}
\begin{equation}\label{Fb}
F^b(0,q)=-2i \pi\alpha_s\int\frac{d^4 P}{(2\pi)^4}(2p_0-k_0)^2
[\bigtriangleup_F(P-K)\bigtriangleup_R(P)+\bigtriangleup_A(P-K)\bigtriangleup_F(P)],
\end{equation}
\begin{equation}\label{Fc}
F^c(0,q)=4i \pi\alpha_s\int\frac{d^4
P}{(2\pi)^4}[\bigtriangleup_R(P)+\bigtriangleup_A(P)+\bigtriangleup_F(P)],
\end{equation}
where $\alpha_s$ is the effective strength for the coupling with
scalar bound states, capital letters denote for four-momentum,
$\bigtriangleup_F$, $\bigtriangleup_R$ and $\bigtriangleup_A$ are
the propagators in Keldysh
representation\cite{Keldysh,Carrington2} which read,
\begin{eqnarray}
\bigtriangleup_{R,A}(P)&=&\frac{1}{P^2-m_s^2\pm i
sgn(p_0)\varepsilon},\nonumber \\[0.3cm]
\bigtriangleup_F(P)&=&-2\pi i[1+n_b(|p_0|)\delta(P^2-m_s^2)],
\end{eqnarray}
where $m_s$ is the scalar mass, $\varepsilon$ is the
infinitesimal, $sgn$ denotes for the sign function and
$n_b(|p_0|)=(e^{\beta |p_0|}-1)^{-1}$ is the Bose-Einstein
distribution function.

Separating the vacuum contribution from Eqs.(\ref{Fb}) and
(\ref{Fc}), one can obtain
\begin{equation}\label{Fbcvac}
F^{b+c}_{vac}(0,q)=- \frac{\alpha_s}{4\pi}q^2\int^1_0 dx (1-x)^2
\log \left (\frac{m_s^2}{m_s^2+x(1-x)q^2}\right),
\end{equation}

\begin{equation}\label{Fbcmatt}
F^{b+c}_{matt}(0,q)=-\frac{4\alpha_s}{\pi}\int^\infty_0 dp\
\frac{p^2}{\omega_s}\left[\frac{\omega_s^2}{p\ q}\log
\left(\frac{q-2p}{q+2p}\right)-1\right]n_b(\omega_s),
\end{equation}
where the vacuum contribution is manipulated in the same
renormalization scheme as Eq.(\ref{Favac}), where
$\omega_s=\sqrt{p^2+m_s^2}$.

\section{Numerical results}

%\subsection{QED case}
To see the oscillatory behavior of the interparticle  potential,
we begin with the pure QED coupling case, i.e., we consider only
Fig.1(a) in $F$. In the following numerical calculations, only
attractive interaction between two fermions with opposite sign is
studied, so that $Q_1 Q_2=-4\pi \alpha$.

\begin{figure}
\begin{minipage}[t]{8cm}
   \resizebox{7cm}{!}{\includegraphics{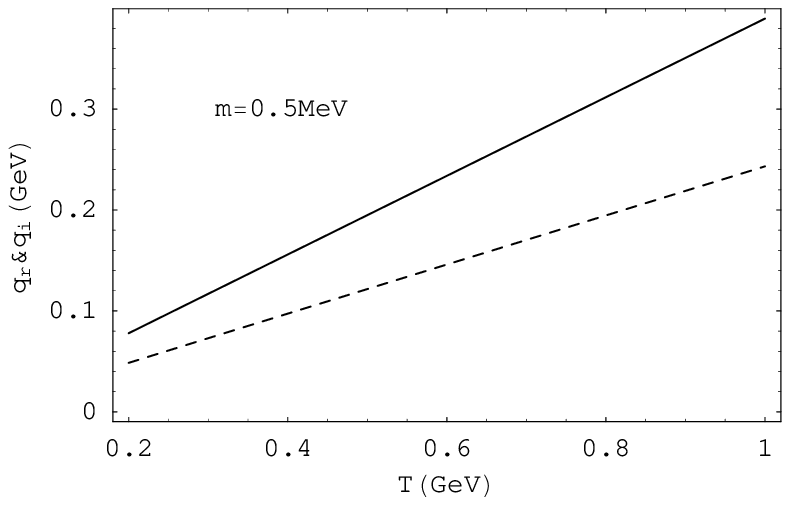}}
 \centering{\par{\hspace{1cm}\scriptsize(a)}}
\end{minipage}
\begin{minipage}[t]{8cm}
   \resizebox{7cm}{!}{\includegraphics{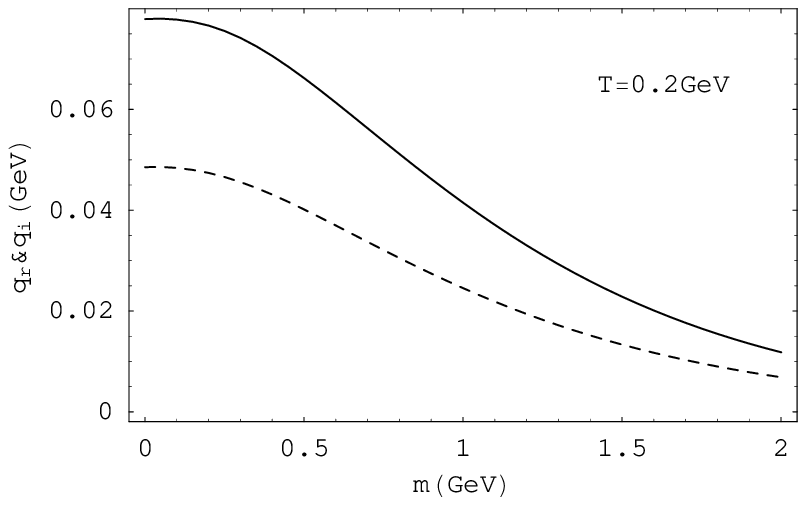}}
 \centering{\par{\hspace{1.2cm}\scriptsize(b)}}
\end{minipage}
\caption{Position of poles in terms of temperature and fermion
mass in pure QED coupling, where the solid lines are for $q_r$ and
the dashed lines are for $q_i$. The curves in (a) are plotted with
fixed fermion mass $m=0.5$MeV and those in (b) are plotted with
fixed temperature $T=0.2$GeV.\label{pole}}
\end{figure}

Inserting the sum of Eqs.(\ref{Favac}) and (\ref{Famatt}) into
Eq.(\ref{potential}), the pole of the integrand in the first
quadrant can be identified numerically by replacing all $q$ as
$q_r+iq_i$ in the denominator and then solving the complex
equation. After that, one can calculate the undetermined
parameters $a$ and $b$ through Eq. (\ref{ab}) and consequently
figure out the potential according to Eq. (\ref{osc potential}).
In Fig.2 we demonstrated the real part $q_r$ (solid) and imaginary
part $q_i$ (dashed) of the pole as functions of temperature and
the fermion mass. In Fig.\ref{pole}(a), the mass has been fixed at
$0.5$MeV and in Fig.\ref{pole}(b) the temperature has been fixed
at $0.2$GeV.
According to Eq.(\ref{osc potential}), the fact  $q_r$ and $q_i$
increase with temperature indicates the oscillation damps faster
and more rapidly with increasing temperature. As to the fermion
mass, there exist an opposite tendency, i.e., fast damping and
rapid oscillations will be achieved by decreasing the mass.

\begin{figure}
\begin{minipage}[t]{8cm}
   \resizebox{7cm}{!}{\includegraphics{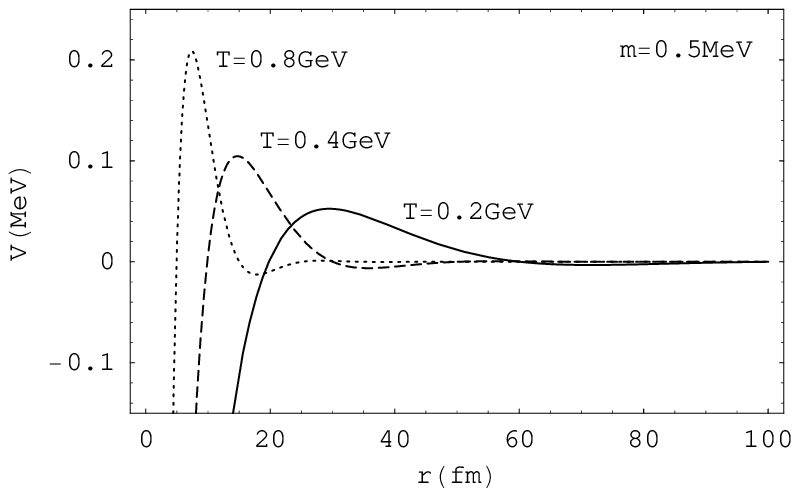}}
 \centering{\par{\hspace{1cm}\scriptsize(a)}}
\end{minipage}
\begin{minipage}[t]{8cm}
   \resizebox{7cm}{!}{\includegraphics{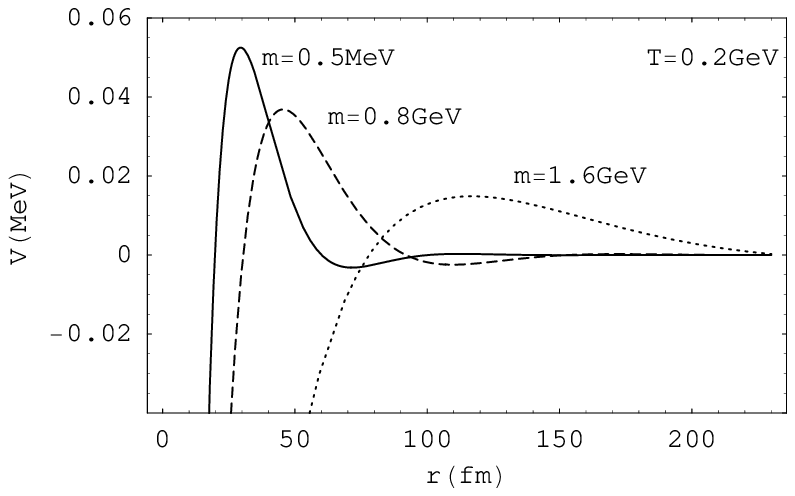}}
 \centering{\par{\hspace{1.2cm}\scriptsize(b)}}
\end{minipage}
\caption{Pure QED interparticle potential oscillates in distance.
The curves in (a) are for fixed mass $m=0.5$MeV with the
temperatures are $0.8$GeV(dotted), $0.4$GeV(dashed) and
$0.2$GeV(solid) respectively. The curves in (b) are for fixed
temperature $T=0.2$GeV with the effective fermion masses are
$0.5$MeV(solid), $0.8$GeV(dashed) and $1.6$GeV(dotted)
respectively.\label{plot_V_QED}}
\end{figure}

To make this analysis more visible, we choose three points on each
curve in Fig.\ref{pole} to plot the oscillatory potential in
Fig.\ref{plot_V_QED},
where  Fig.\ref{plot_V_QED}(a) is  fixed at $m=0.5$MeV with the
temperatures are $0.8$GeV(dotted), $0.4$GeV(dashed) and
$0.2$GeV(solid) respectively. Fig.\ref{plot_V_QED}(b) is fixed at
$T=0.2$GeV with the effective fermion masses are $0.5$MeV(solid),
$0.8$GeV(dashed) and $1.6$GeV(dotted) respectively. Note that the
solid curve in the two plots share the same set of parameters, and
thus can be chosen as a reference.

In Fig.\ref{plot_V_QED}, one can see clearly the interparticle
potential oscillates in distance, where the peaks and the
oscillating damping tail vary with temperature and mass. The first
peak, which is the most obvious one on the plot and can be
regarded as the representation of all the peaks, decreases in
amplitude and becomes broad in width with increasing temperature,
which means the effective interacting distance becomes long though
weak at relatively low temperature. Contrary to the temperature
effect, the decreasing fermion mass enhances and narrows the peak
as shown in Fig.\ref{plot_V_QED}(b).

\begin{figure}
\begin{minipage}[t]{8cm}
   \resizebox{7cm}{!}{\includegraphics{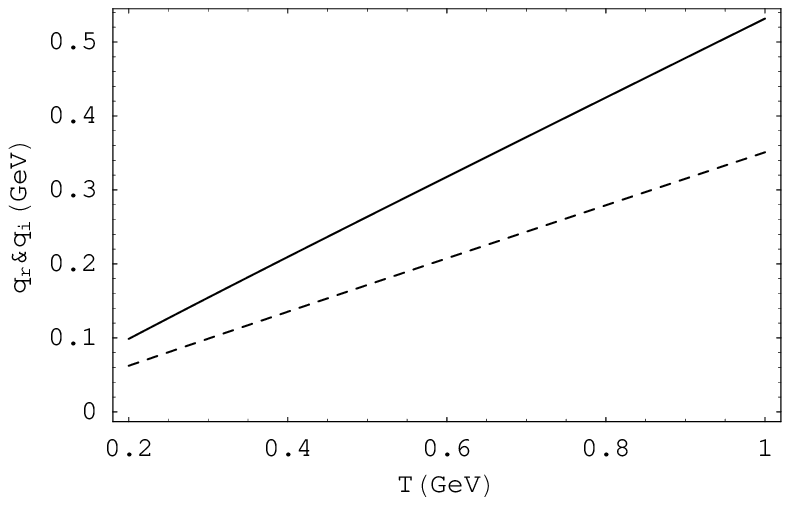}}
\centering{\par{\hspace{1cm}\scriptsize{$\alpha=\frac{1}{137}$,
$\alpha_s=0.02$, $m_s=0.4$GeV, $m=10$MeV}\\
\hspace{1cm}\scriptsize(a)}}
\end{minipage}
\begin{minipage}[t]{8cm}
   \resizebox{7cm}{!}{\includegraphics{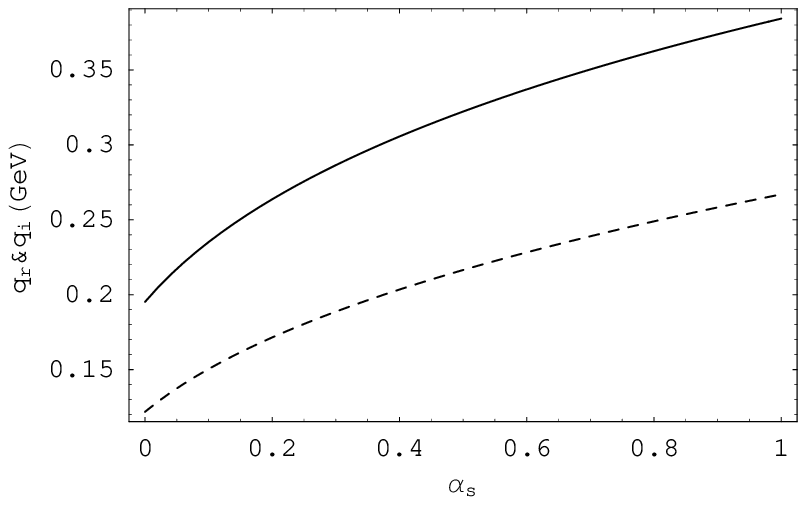}}
 \centering{\par{\hspace{1cm}\scriptsize{$\alpha=\frac{1}{137}$,
 $T=0.5$GeV,
$m_s=0.4$GeV, $m=10$MeV}\\ \hspace{1cm}\scriptsize(b)}}
\end{minipage}\\[1cm]
\begin{minipage}[t]{8cm}
   \resizebox{7cm}{!}{\includegraphics{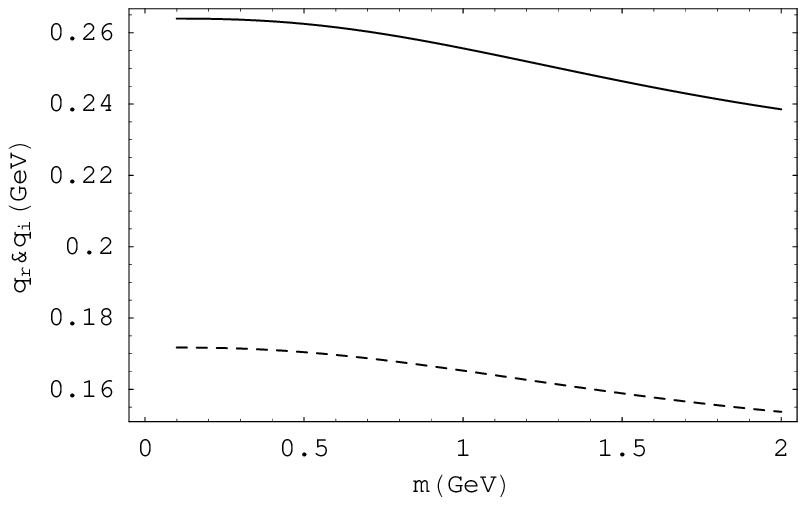}}
 \centering{\par{\hspace{1cm}\scriptsize{$\alpha=\frac{1}{137}$, $\alpha_s=0.02$,
$m_s=0.4$GeV, $T=0.5$GeV}\\ \hspace{1cm}\scriptsize(c)}}
\end{minipage}
\begin{minipage}[t]{8cm}
   \resizebox{7cm}{!}{\includegraphics{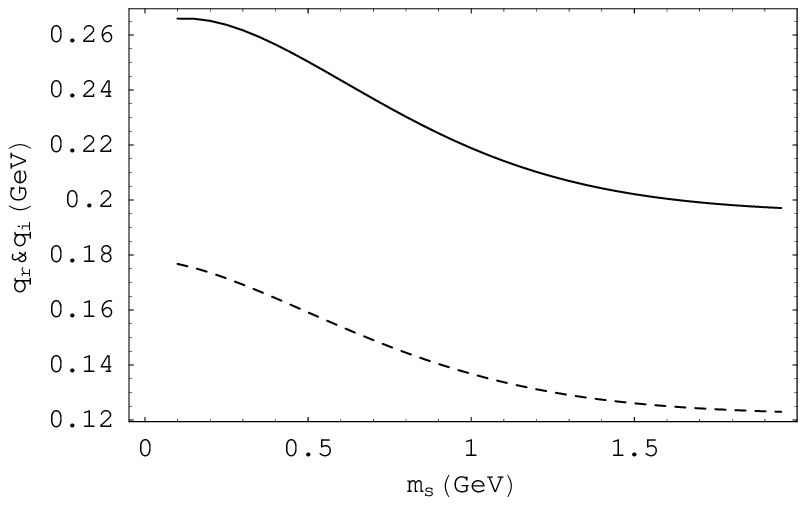}}
 \centering{\par{\hspace{1cm}\scriptsize{$\alpha=\frac{1}{137}$, $\alpha_s=0.02$,
$T=0.5$GeV, $m=10$MeV}\\ \hspace{1cm}\scriptsize(d)}}
\end{minipage}
\caption{Position of poles in terms of $T$, $\alpha_s$, $m$ and
$m_s$, where the solid lines are for $q_r$ and the dashed lines
for $q_i$ as in Fig.\ref{pole}. The parameters we used in the
calculation are presented below each plot. \label{pole_sQED}}
\end{figure}

\begin{figure}
%\begin{minipage}[t]{0.5\linewidth}
 %\centering
 %\begin{center}
   \resizebox{8cm}{!}{\includegraphics{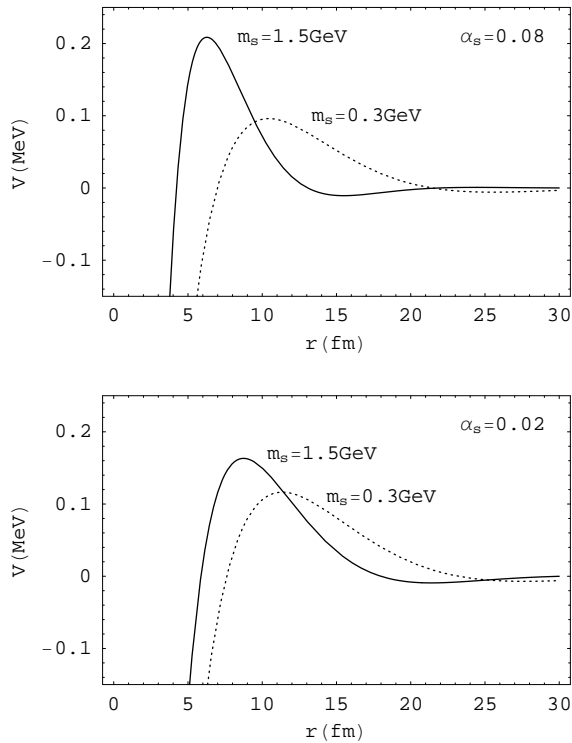}}
 %\end{center}
\caption{In-medium interparticle potential varies with the scalar
mass at different $\alpha_s$. The solid line stands for the case
of $m_s=0.5$GeV and the dotted line for the heavier mass case of
$m_s=1.5$GeV. The difference between the upper and the lower
frames lies in $\alpha_s$, where the former is for $\alpha_s=0.02$
and the latter is for $\alpha_s=0.08$.\label{Vms_c_sQED}}
%\end{minipage}%
\end{figure}

%\subsection{QED+sQED coupling}
Now we are in the position to present the effect of bound states.
To involve this effect, the contributions from the last two
diagrams in Fig.1 should be added to $F$, i.e.,
\begin{equation}
F=F^a_{vac}+F^a_{matt}+F^{b+c}_{vac}+F^{b+c}_{matt}.
\end{equation}
Numerically figuring out the zero point of $q^2+F$ in
Eq.(\ref{potential}), one can obtain the oscillatory potential
with respect to different group of parameters. In
Fig.\ref{pole_sQED}, we exhibit $q_r$(solid) and $q_i$(dashed) in
terms of $T$, $\alpha_s$, $m$ and $m_s$. The parameters we used in
the calculation are presented below each plot.
As a general impression, one can find out two tendencies: one is
ascending with increasing parameters, like the first row in
Fig.\ref{pole_sQED}.  The other is descending with increasing
parameters, like the second row. Based on the previous experience
on pure QED coupling, one could expect the first tendency would
enhance but narrow the peaks of interparticle potential and shift
them to the left, which means strong but short distance
interaction. While the second tendency would suppress but broaden
the peaks and shift them to the right, which suggests the weak but
long distance interaction. We show these tendencies more clearly
in Fig.\ref{Vms_c_sQED} by directly exhibiting the potential in
terms of these parameters where we set: $\alpha=\frac{1}{137}$,
$m=10$MeV and $T=0.5$GeV.

One can observe that Fig.\ref{Vms_c_sQED} is consistent with what
we have argued by just analyzing the evolutional tendencies of the
pole position with respect to various parameters. In each plot of
Fig.\ref{Vms_c_sQED}, the two curves denote for the case of
$m_s=1.5$GeV(solid) and $m_s=0.3$GeV(dashed) respectively, which
shows the decreasing of $m_s$ or increasing of $\alpha_s$ enhances
but narrows the peaks on the potential and shift them to the left,
indicating the strong but short distance interaction. Besides, it
is worthwhile to compare the differences between the two curves in
the same plot from the upper and lower ones which are mapped in
different $\alpha_s$. The comparison shows the stronger the
coupling $\alpha_s$ is, the more the scalar mass can affect the
shape of the potential, even the other parameters are the same.

\begin{figure}
%\begin{minipage}[t]{0.5\linewidth}
 %\centering
 %\begin{center}
   \resizebox{8cm}{!}{\includegraphics{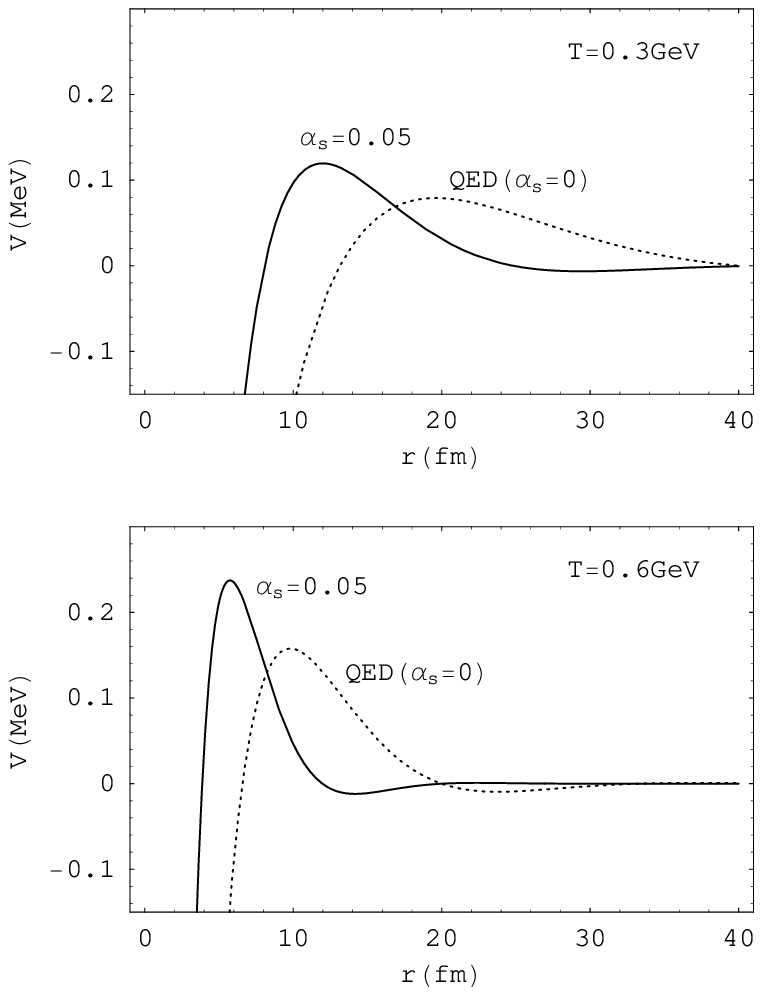}}
 %\end{center}
\caption{Interparticle potential of QED and QED+sQED. The upper
plot is at $T=0.3$GeV and the lower plot is at $T=0.6$GeV. The
dotted curves in both plots are for the pure QED potential and the
solid curves are for the potential involving scalar bound states
where the coupling strength
$\alpha_s=0.05$.\label{Vc_t_sQEDvsQED}}
%\end{minipage}%
\end{figure}

%\subsection{QED vs. QED+sQED}
Finally let us compare the pure QED with the case involving scalar
bound states. In Fig.\ref{Vc_t_sQEDvsQED}, we demonstrated the
potential at $\alpha_s=0$ (pure QED) and $\alpha_s=0.05$ in upper
plot with $T=0.3$GeV and arrayed it with the lower one at
$T=0.6$GeV. The masses of the fermion and bound state are 10MeV
and 0.4GeV respectively. There is no surprise that the pure QED
potential is longer in the interactive range but weaker in
amplitude than that involves the coupling with scalar bound state
since it is the limit of the small coupling strength.

\iffalse %
Finally let us compare the effective interactive distance
of oscillatory potential with that of the Debye potential. In
Debye screening,
\begin{equation}
V_D(r)=\frac{Q_1 Q_2}{4\pi}\frac{e^{-m_Dr}}{r},
\end{equation}
where Debye mass is defined as the static self-energy
$m^2_D=\Pi_{00}^{\mbox{\tiny{HTL}}}(q_0=0,q)$ at HTL
approximation. For QED, it is $m^2_D=\frac{4\pi}{3}\alpha T^2$
\cite{LeBellac}. At $T=0.5$GeV, the effective interactive range,
namely Debye radius, is the reciprocal of Debye mass which is
around $2.3$ fm. This value is much smaller than that of the
oscillatory potential referring to the solid curve in the lower
plot of Fig.\ref{Vms_c_sQED}.

\fi

\section{Summary and discussions}

We investigated the in-medium interparticle potential of gauge
plasma with bound states by employing QED and sQED coupling, where
complete one-loop boson self-energy has been involved which is
equivalent to resum all one-loop diagrams. Different from Debye
screening via HTL approximation, an oscillatory behavior of the
interparticle potential with exponential damping amplitude has
been demonstrated which is referred to the so-called Yukawa
oscillation. The four parameters, $\alpha_s$, $T$, $m$, and $m_s$
in our model are free to change, since we do not intend to study
the electromagnetic system but to employ it as a toy coupling
form. In the numerical camputation we fixed $\alpha$ to the fine
structure constant by considering the matter particle as
point-like. While $\alpha_s$ was treated as a effective parameter
because of the inner structure of the scalar bound state.
Evidently, $\alpha_s=0$ recovers the pure QED coupling. In the
evaluation of the integral in Eq.(\ref{potential}), we first found
out the complex poles of the integrand, tracking their steps with
parameters evolution. Then two opposite tendencies of the
potential variation in terms of different parameters have been
reported. The first tendency is to suppress the potential
amplitude, slow down the oscillation "frequencies" and shift the
peaks' centers to long distance. This tendency is caused by either
decreasing $\alpha_s$ and $T$ or increasing $m$ and $m_s$.  The
second opposite tendency is to enhance the potential amplitude and
frequencies, shifting the peaks center to short distance. As
expected, this tendency is due to the increase of $\alpha_s$ and
$T$ or decrease of $m$ and $m_s$. In brief, the potential between
two fermions can be either long-weak or short-strong, according to
the competition among the two groups of parameters. In some
circumstances when the parameters are carefully ''tuned'', this
potential is on the chance of being qualified strong and long
enough to produce liquid state.

We would like to point out here that the behaviors of the
potential are governed by the static modes which are determined by
the dispersion relation of plasma. The Debye screening, which
describes the high temperature environment, can be obtained in the
HTL approximation to the boson self-energy where the external
momenta are soft. This implies only soft modes at $q\sim gT$ get
involved in the dielectric function when evaluating the
interparticle potential. Therefore only those static modes in the
soft region could be selected via dispersion relation and finally
promise Debye screening. In other words, to reach Debye screening,
one should take the high temperature limit by employing the HTL
approximation from the beginning in the dielectric function to
separate the soft modes, but not take the high temperature limit
of $q_r$ and $q_i$ in Eq. (\ref{osc potential}) directly. As a
matter of fact, the HTL static modes which are adapted to describe
the extremely high temperature system may lose some significant
details (hard mode contributions) in describing the medium at
$1\sim 3$ times critical temperature. Instead, the improved
calculation in this paper picks up the the long distance
oscillation tail of the potential which might be helpful to
understand the low viscous mechanism of QGP with scalar bound
state.

The interparticle potential is physically measurable at least in
principle so we expect it is gauge independent. As to QED
calculation, gauge problem does not emerge because the boson
self-energy is gauge independent. While for QCD, the linear
framework can only be persisted in the temporal axis gauge (TAG),
but the one-loop calculations involving gluon self-coupling are
various in different gauges even to the static modes. Anyway, we
hope the discussion in TAG may give the qualitative features as
well as correct directions of their evolution. More detailed
investigations such as resummations are desirable in the future.

\begin{acknowledgments}
This work is partly supported by the National Natural Science
Foundation of China under project Nos. 10675052 and 10575043, the
Ministry of Education of China with Project No. CFKSTIP-704035 and
NCET-05-0675.
\end{acknowledgments}

 \end{document}